\documentclass[aps,prd,reprint,onecolumn,notitlepage,superscriptaddress,nofootinbib]{revtex4-2}
\usepackage{amssymb,amsthm,amsmath}
\usepackage[utf8]{inputenc}
\usepackage{textcomp}
\usepackage{slashed}
\usepackage{verbatim}
\usepackage{braket}
\usepackage{mathrsfs}
\usepackage{graphicx}
\usepackage{booktabs}
\usepackage{rotating}
\usepackage{multirow}
\usepackage[caption=false]{subfig}
\usepackage[hidelinks]{hyperref}

\newcommand{\SU}{\mathrm{SU}}
\newcommand{\U}{\mathrm{U}}

\begin{document}
\title{ Nonstandard Solution for Anomaly Cancellation \\
as Seesaw Neutrino Origin in the SM}
\author {Zi-Yue Zou}\email{ziy\_zou@sjtu.edu.cn}
\affiliation{State Key Laboratory of Dark Matter Physics, Tsung-Dao Lee
Institute \& School of Physics and Astronomy, Shanghai Jiao Tong University,
Shanghai 200240, China}
\affiliation{Key Laboratory for Particle Astrophysics and Cosmology (MOE) \&
Shanghai Key Laboratory for Particle Physics and Cosmology, Tsung-Dao Lee
Institute  \& School of Physics and Astronomy, Shanghai Jiao Tong University,
Shanghai 200240, China}
\author {Chia-Wei Liu}\email{chiaweiliu@ucas.ac.cn}
\affiliation{School of Fundamental Physics and Mathematical Sciences, Hangzhou
Institute for Advanced Study, UCAS, Hangzhou 310024, China}
\author { Zhong-Lv Huang}\email{huangzhonglv@sjtu.edu.cn}
\affiliation{State Key Laboratory of Dark Matter Physics, Tsung-Dao Lee
Institute \& School of Physics and Astronomy, Shanghai Jiao Tong University,
Shanghai 200240, China}
\affiliation{Key Laboratory for Particle Astrophysics and Cosmology (MOE) \&
Shanghai Key Laboratory for Particle Physics and Cosmology, Tsung-Dao Lee
Institute  \& School of Physics and Astronomy, Shanghai Jiao Tong University,
Shanghai 200240, China}
\author {Xiao-Gang He}\email{hexg@sjtu.edu.cn}
\affiliation{State Key Laboratory of Dark Matter Physics, Tsung-Dao Lee
Institute \& School of Physics and Astronomy, Shanghai Jiao Tong University,
Shanghai 200240, China}
\affiliation{Key Laboratory for Particle Astrophysics and Cosmology (MOE) \&
Shanghai Key Laboratory for Particle Physics and Cosmology, Tsung-Dao Lee
Institute  \& School of Physics and Astronomy, Shanghai Jiao Tong University,
Shanghai 200240, China}
\date{November 3, 2026}

\begin{abstract}
For fixed Standard Model (SM) non-Abelian representations of 15 chiral fermions
with arbitrary hypercharges, anomaly cancellation admits the usual assignment
and a distinct nonstandard solution.
In the latter, the exotic quark and lepton weak doublets and exotic lepton
singlet have zero hypercharge, whereas the two exotic quark singlets carry
opposite hypercharges $-q$ and $q$.
The neutral exotic lepton singlets naturally provide heavy neutrinos. With two
copies, the minimal model realizes the minimal type-I seesaw, yielding one
massless active neutrino, and predicts
$m_{\beta\beta}=1.2\text{--}4.1\,\mathrm{meV}$ for normal ordering and
$15.9\text{--}48.9\,\mathrm{meV}$ for inverted ordering.
In a direct SM realization, generating exotic-quark masses through the SM Higgs
mechanism fixes their electric charges to $\pm 1/2$.
A separate $\SU(2)_{L'}$ realization of the nonstandard solution can allow
exotic-quark masses ranging from several TeV to $10\,\mathrm{TeV}$ with
order-one Yukawa couplings while all charged exotic particles carry charges $\pm
q$. In both cases, the lightest exotic quark and lepton are generically stable,
providing distinctive and potentially testable signatures while remaining
compatible with current constraints.
\end{abstract}

\maketitle

\noindent{\bf Introduction.} The Standard Model (SM) is the most successful
model in particle physics. Quantum anomalies play a central role in modern
particle physics and gauge anomaly cancellation is one of the basic consistency
requirements of a chiral gauge theory such as the SM. It is well known that a
symmetry of the classical action need not survive quantization, as exemplified
by the axial anomaly ~\cite{Adler:1969gk,Bell:1969ts}, while an uncanceled gauge
anomaly would destroy the consistency of a gauge
theory~\cite{Bouchiat:1972iq,tHooft:1971qjg,Gross:1972pv}. It is a remarkable
fact that gauge anomaly cancellation leads quarks and leptons to have particular
fractional hypercharges, thereby yielding the observed quantization of electric
charge. This was first pointed out in Ref.~\cite{Geng:1989tcu} and subsequently
studied in Refs.~\cite{Babu:1989ex,Foot:1992ui}.
It is not widely known that the same anomaly-cancellation equations also admit
the nonstandard solution discussed in Refs.~\cite{Minahan:1989vd,Geng:1990nh,
He:1990me}, which leads to exotic new particles.
The implications of the nonstandard solution remain to be established.
A natural question is whether this exotic solution can provide a viable
additional fermion sector with distinctive phenomenological consequences.
Here we show that this solution provides a natural origin for the minimal type-I
seesaw, yielding one massless active neutrino.

A standard approach to charge quantization in the SM is to impose the
anomaly-cancellation conditions on a single SM-like generation of 15 Weyl
fermions with fixed $\SU(3)_C\times\SU(2)_L$ representations and arbitrary
hypercharges.
Besides the standard solution with the SM hypercharge assignments, the anomaly
equations admit a nonstandard solution with exotic hypercharges. This solution
contains an exotic quark weak doublet, an exotic lepton weak doublet, and an
exotic lepton singlet, all with zero hypercharge, while the two exotic quark
singlets carry opposite hypercharges $-q$ and $q$.
The nonstandard solution has several notable features. All exotic-lepton
doublets can acquire nonzero masses only if the number of exotic families is
even. The minimal choice of two families then provides two neutral singlets that
can serve as heavy neutrinos, naturally realizing the minimal type-I seesaw with
one massless active neutrino. The exotic-quark sector, however, presents
phenomenological challenges because the lightest state is stable and may carry
fractional electric charge, while sufficiently large masses in a direct SM
realization require strong Yukawa couplings and are difficult to reconcile with
experimental constraints.
Finally, we consider an extension with a separate $\SU(2)_{L'}$ gauge group.
This extension shifts the exotic-particle masses to a higher symmetry-breaking
scale, allowing the experimental constraints to be satisfied.
\\

\noindent{\bf Algebraic solutions of the anomaly equations.} The SM gauge group
is
$G_{\rm SM}=\SU(3)_C\times \SU(2)_L\times \U(1)_Y$. We write all chiral fermions
as left-handed Weyl fields.
A generic field is denoted by
\begin{eqnarray}
\psi_i \sim (\mathbf{R}_{3i},\mathbf{R}_{2i},y_i),
\end{eqnarray}
where $\mathbf{R}_{3i}$ and $\mathbf{R}_{2i}$ are representations of $\SU(3)_C$
and $\SU(2)_L$, and $y_i$ is its hypercharge. With this convention, one SM
family consists of five gauge multiplets containing a total of 15 left-handed
Weyl fermion degrees of freedom
\begin{equation}
\label{sm_sol}
Q\sim({\mathbf{3},\mathbf{2}},1/6),\quad
u^c\sim({\overline{\mathbf{3}},\mathbf{1}},-2/3), \quad 
d^c\sim({\overline{\mathbf{3}},\mathbf{1}},1/3),\quad
L\sim({\mathbf{1},\mathbf{2}},-1/2), \quad
e^c\sim({\mathbf{1},\mathbf{1}},1).
\end{equation}
The superscript $c$ denotes charge conjugation, $X^c\equiv C\overline{X}^{\,T}$.
Thus, $u^c$, $d^c$, and $e^c$ are the left-handed charge conjugates of the usual
right-handed SM fields.

To determine whether this pattern is unique, consider the same 15 fermion
representations as a single SM family but leave the hypercharges $y_i$ arbitrary
as follows:
\begin{equation}
Q\sim({\mathbf{3},\mathbf{2}},y_1),\quad 
u^c\sim({\overline{\mathbf{3}},\mathbf{1}},y_2),\quad
d^c\sim({\overline{\mathbf{3}},\mathbf{1}},y_3),\quad 
L\sim({\mathbf{1},\mathbf{2}},y_4),\quad
e^c\sim({\mathbf{1},\mathbf{1}},y_5).
\label{eq:general_family}
\end{equation}
The two color anti-triplet singlets have identical non-Abelian quantum numbers,
so $y_2\leftrightarrow y_3$ only relabels them. We remove this redundancy from
the outset by taking $y_3\geq y_2$.

There are two types of potential anomalies to consider for the fermion content
above. The first is the global Witten $\SU(2)$ anomaly, whose cancellation
requires an even number of $\SU(2)$ fundamental representations. This condition
is satisfied by the 15 Weyl fermions because the number of left-handed weak
doublets, including the color multiplicity of $Q$, is
$3+1=4$~\cite{Witten:1982fp}.

The second type comprises perturbative triangle anomalies. We also include the
mixed gauge--gravity anomaly~\cite{Delbourgo:1972xb,Alvarez-Gaume:1983ihn},
whose coefficient is the sum of $y_i$ over all 15 fermion degrees of freedom.
The nontrivial combinations are $\mathcal A_{33Y}=[\SU(3)_C]^2\U(1)_Y$,
$\mathcal A_{22Y}=[\SU(2)_L]^2\U(1)_Y$, $\mathcal A_{YYY}=[\U(1)_Y]^3$, and the
mixed gauge--gravity anomaly $\mathcal A_{\mathrm{grav}^2Y}$. Their cancellation
gives
\begin{equation}
\begin{aligned}
\mathcal A_{33Y}&:\quad 2y_1+y_2+y_3=0,\\
\mathcal A_{22Y}&:\quad 3y_1+y_4=0,\\
\mathcal A_{YYY}&:\quad 6y_1^3+3y_2^3+3y_3^3+2y_4^3+y_5^3=0,\\
\mathcal A_{\mathrm{grav}^2Y}&:\quad 6y_1+3y_2+3y_3+2y_4+y_5=0.
\end{aligned}
\label{eq:anomaly_conditions}
\end{equation}
Eliminating $y_3$, $y_4$, and $y_5$ in favor of $y_1$ and $y_2$ using the three
equations linear in the $y_i$, ${\cal A}_{YYY}$ gives
\begin{eqnarray}
-18y_1\,(y_2+4y_1)(y_2-2y_1)=0.
\label{eq:cubic_factorized}
\end{eqnarray}

The independent anomaly-free solutions are therefore
\begin{equation}
\begin{aligned}
\text{SM:}\quad& y_2=-4y_1,\quad y_3=2y_1,\quad
 y_4=-3y_1,\quad y_5=6y_1,\quad y_1\geq0,\\
\text{exotic:}\quad& y_1=y_4=y_5=0,\quad y_2=-q,\quad y_3=q,\quad q\geq0.
\label{eq:anomaly_solution}
\end{aligned}
\end{equation}
The SM uses the normalization of  $y_5=1$ in Eq.~\eqref{sm_sol}~\footnote{The
anomaly equations fix only hypercharge ratios. For the SM Higgs doublet
$\Phi\sim(\mathbf{1},\mathbf{2},y_\Phi)$, neutrality of the vacuum expectation
value (VEV) $\langle\Phi\rangle=(0,v/\sqrt{2})^T$ under $Q_{\rm em}=T_3+Y$ fixes
$y_\Phi=1/2$, while gauge invariance of $U^c\Phi Q$, together with $y_2=-4y_1$,
gives $y_1=1/6$.}, while in the usual solution neutrinos cannot have a
renormalizable mass term.
The exotic solution is qualitatively different and gives
\begin{equation}
Q' \sim (\mathbf{3},\mathbf{2},0),\quad 
U^c \sim (\overline{\mathbf{3}},\mathbf{1},-q),\quad
D^c \sim (\overline{\mathbf{3}},\mathbf{1},q),\quad 
E\sim (\mathbf{1},\mathbf{2},0),\quad
N \sim (\mathbf{1},\mathbf{1},0).
\label{eq:exotic_solution}
\end{equation}
Compared with the SM quark doublet, the exotic quark doublet $Q'$ may have
either chirality: it can be realized as left-handed or right-handed, provided
the singlet fields are conjugated consistently.
Unlike a sequential SM family, the exotic quark weak doublet, the lepton weak
doublet, and the singlet all have zero hypercharge, while $U^c$ and $D^c$ carry
$-q$ and $q$, respectively. We adopt the convention $q >0$, which fixes the
otherwise interchangeable labels of the two singlets. Once the SM hypercharge
normalization is fixed, anomaly cancellation leaves the value of $q$
undetermined.
The exotic lepton singlet $N$ is particularly interesting because it can be the
origin of neutrino mass generation through the seesaw mechanism.
\\

\noindent{\bf The exotic solution and the minimal seesaw mechanism.}
We now embed the exotic solution in Eq.~\eqref{eq:exotic_solution} into the SM,
together with the usual anomaly-free fermion content of its three generations,
and examine whether it can be implemented consistently at the phenomenological
level, particularly the generation of exotic-particle masses.

The masses in the exotic lepton sector arising from the nonstandard solution
\begin{eqnarray}
E=(E_{+1/2},E_{-1/2})^T\sim({\mathbf{1},\mathbf{2}},0),\qquad
N\sim({\mathbf{1},\mathbf{1}},0),
\end{eqnarray}
need to be carefully treated. Although $N$ can have a Majorana mass through the
bare mass term $M_N\overline{N^c}N$, $E$ cannot have a bare mass if there is
only one generation of this exotic family.
The exotic lepton doublets \(E_i\) can acquire masses when more than one exotic
generation is introduced:
\begin{equation}
\mathcal{L}\supset
-(M_E)_{ij}\epsilon_{ab}\overline{(E_i^a)^c}E_j^b
\,,
\label{eq:lepton_antisym_mass}
\end{equation}
where \(a,b=1,2\) are \(\SU(2)_L\) indices, while \(i,j=1,\ldots,n\) label the
\(n\) copies of the exotic family, and \(\epsilon_{12}=+1\). Since
\(M_E^T=-M_E\), we have \(\det(M_E)=0\) for any odd number of generations,
implying that at least one exotic lepton doublet \(E\) remains massless. Thus,
two complete copies of the exotic family constitute the minimal viable
realization of this mass-generation mechanism.
Interestingly, the presence of two copies can realize the minimal type-I seesaw
mechanism, generating masses for the SM neutrinos:
\begin{equation}
\mathcal{L}\supset
-\frac{1}{2}(M_N)_{ij}\overline{(N_i)^c}N_j
-Y_{\ell i}\epsilon_{ab}\overline{(L_\ell^a)^c}\Phi^bN_i
+\mathrm{h.c.},
\label{eq:seesaw_lagrangian}
\end{equation}
where \(\ell=e,\mu,\tau\) labels the SM lepton flavors. Therefore, \(Y_{\ell
i}\) is a \(3\times2\) Yukawa matrix, while \((M_N)_{ij}\) is a \(2\times2\)
symmetric Majorana mass matrix.

In the seesaw limit, the leading light-neutrino mass matrix is
\begin{eqnarray}
 m_\nu\simeq -
 \frac{v^2}{2}
Y  M_N^{-1} Y^T .
\label{eq:type_i_seesaw}
\end{eqnarray}
This is a particular case of the type-I seesaw
mechanism~\cite{Minkowski:1977sc,Yanagida:1979as,Gell-Mann:1979vob,Mohapatra:1979ia}.
The rank of $m_\nu$ is at most two, so one active neutrino is massless at
leading order~\cite{Frampton:2002qc,Ibarra:2003up,Davidson:2006tg,Dev:2026ddq}.
Consequently, once the neutrino mass ordering is specified, the two nonzero
neutrino masses are fixed by using the current information on $\Delta
m^2_{ij}$~\cite{KamLAND-Zen:2024eml,JUNO:2025gmd,Esteban:2026phq}.  Adopting the
best-fit results and the corresponding $1\sigma$ uncertainties from the
NuFIT~6.1 global analysis ~\cite{Esteban:2024eli,NuFIT:6.1},
we obtain
\begin{align}
&\text{Normal ordering\,(NO):}\quad 
    m_1=0,\qquad
    m_2=8.68^{+0.05}_{-0.06}~\mathrm{meV},
    \quad 
    m_3=50.11^{+0.21}_{-0.20}~\mathrm{meV},
    \label{eq:nu-mass-NO}
    \\
    &\text{Inverted ordering\,(IO):}\quad 
    m_3=0,\qquad
    m_1=49.07^{+0.20}_{-0.20}~\mathrm{meV},
    \quad 
    m_2=49.83^{+0.20}_{-0.20}~\mathrm{meV}.
    \label{eq:nu-mass-IO}
\end{align}

In this rank-two limit, we now examine the effective mass $m_{\beta\beta}$
governing neutrinoless double-beta $(0\nu\beta\beta)$
decay~\cite{Furry:1939qr,Schechter:1981bd}.
We assume that     $0\nu\beta\beta$ decay    is dominated by light Majorana
neutrino exchange.
Using the Particle Data Group convention~\cite{ParticleDataGroup:2026mpi},
one obtains
\begin{equation}
m_{\beta\beta}=
\left|m_1c_{12}^{2}c_{13}^{2}
+m_2s_{12}^{2}c_{13}^{2}e^{i\alpha_{21}}
+m_3s_{13}^{2}e^{i(\alpha_{31}-2\delta)}\right|,
\end{equation}
where $s_{ij}\equiv\sin\theta_{ij}$ and $c_{ij}\equiv\cos\theta_{ij}$.  Although
the construction contains two sterile Majorana fermions $N_{1,2}$, it does not
give two independent low-energy Majorana phases in $m_{\beta\beta}$.  Since one
light-neutrino eigenstate is massless, it can be rephased, and only one physical
relative phase remains.  The two orderings therefore reduce to
\begin{align}
m_{\beta\beta}^{\rm NO}
&=\left|
\sqrt{\Delta m_{21}^{2}}\,s_{12}^{2}c_{13}^{2}
+\sqrt{\Delta m_{31}^{2}}\,s_{13}^{2}e^{i\varphi_{\rm NO}}
\right|,
\qquad \varphi_{\rm NO}\equiv\alpha_{31}-\alpha_{21}-2\delta,
\nonumber
\\
m_{\beta\beta}^{\rm IO}
&=c_{13}^{2}\left|
\sqrt{|\Delta m_{32}^{2}|-\Delta m_{21}^{2}}\,c_{12}^{2}
+\sqrt{|\Delta m_{32}^{2}|}\,s_{12}^{2}e^{i\varphi_{\rm IO}}
\right|,
\qquad \varphi_{\rm IO}\equiv\alpha_{21}.
\end{align}
Varying $\sin^2\theta_{12}$, $\sin^2\theta_{13}, \Delta m_{21}^{2}$, and
$|\Delta m_{3\ell}^{2}|$ independently over  $3\sigma$ intervals gives the
envelopes
\begin{equation}
m_{\beta\beta}^{\rm NO}=1.17\; \text{--}\; 4.07~{\rm meV},
\qquad
m_{\beta\beta}^{\rm IO}=15.85\; \text{--}\; 48.94~{\rm meV}.
\end{equation}

The $m_{\beta\beta}$ ranges quoted above do not incorporate correlations among
the oscillation parameters. The phase dependence and the resulting uncertainty
bands are shown in Fig.~\ref{fig:mee_majorana_phase}.
The gray hatched region shows the current
KamLAND-Zen limit. The complete data set gives $T_{1/2}^{0\nu}(^{136}{\rm
Xe})>3.8\times10^{26}~{\rm yr}$ at $90\%$ C.L., and it implies an upper limit on
\(m_{\beta\beta}\) ranging from \(28\) to \(122~{\rm meV}\), depending on the
nuclear matrix element used~\cite{KamLAND-Zen:2024eml}.
The current KamLAND-Zen result may already exclude the upper part of the IO band
in Fig.~\ref{fig:mee_majorana_phase}, although this conclusion depends on the
nuclear matrix element adopted.
For comparison, KamLAND2-Zen, a planned upgrade of KamLAND-Zen, aims to reach a
sensitivity of $m_{\beta\beta}<20~{\rm meV}$~\cite{Nakamura:2020szx}.
If no signal is observed at this sensitivity, KamLAND2-Zen will exclude the IO
region with $m_{\beta\beta}\gtrsim20~{\rm meV}$, leaving mainly the region near
$\varphi\simeq\pi$ untested, while the few-meV NO region will remain beyond
reach.
\\

\begingroup
\makeatletter
\renewcommand{\fnum@figure}{\figurename~\thefigure}
\makeatother
\begin{figure}[t]
\centering
\includegraphics[width=0.8\textwidth]{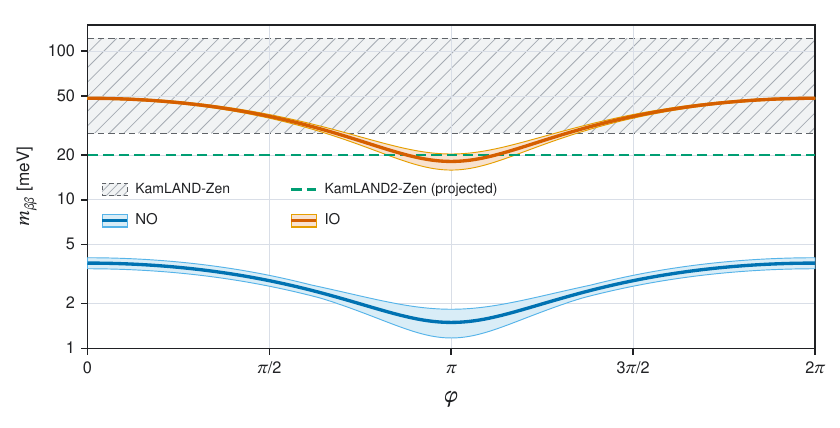}
\caption{
Effective Majorana mass $m_{\beta\beta}$ versus the physical phase $\varphi$ in
the rank-two minimal seesaw for NO (blue) and IO (vermilion). The solid curves
use the NuFIT~6.1 best-fit parameters, while the bands show independent
$3\sigma$ variations without parameter correlations. Gray hatching marks the
current KamLAND-Zen $90\%$ C.L. upper limits, which range from $28$ to $122~{\rm
meV}$ for different nuclear matrix elements. The green dashed line indicates the
projected KamLAND2-Zen sensitivity.
}
\label{fig:mee_majorana_phase}
\end{figure}
\endgroup

\noindent{\bf Problems with exotic quarks and ways out.}
We now examine mass generation for the exotic quarks.
Gauge invariance forbids bare Dirac mass terms for the exotic quarks. For
$q=1/2$, however, their masses can be generated through Yukawa couplings to the
SM Higgs doublet,
\begin{equation}
\mathcal L_q=-y^U_{ij} \overline Q'_i {\widetilde{\Phi}} U_j-y^D_{ij} \overline
Q'_i \Phi D_j+\mathrm{h.c.},\quad 
\widetilde{\Phi}=i\sigma_2\Phi^\ast .
\label{eq:exotic_quark_yukawa}
\end{equation}
In this case, $U^c$ and $D^c$ carry electric charges $-1/2$ and $+1/2$,
respectively. The exotic-quark masses therefore arise after electroweak symmetry
breaking and they can be parametrized in terms of the corresponding Yukawa
couplings as
\begin{eqnarray}
m_U=\frac{y_U v}{\sqrt{2}},\qquad
m_D=\frac{y_D v}{\sqrt{2}} .
\label{eq:exotic_quark_mass}
\end{eqnarray}
The VEV $v$ sets the weak gauge boson masses and is most precisely determined
from the Fermi constant~\cite{ParticleDataGroup:2026mpi},
$v^2=(\sqrt{2}G_F)^{-1}\simeq (246~{\rm GeV})^2$.
Direct searches for heavy exotic quarks already probe the TeV region in standard
benchmark channels~\cite{Jager:2018ecz,CMS:2024eyx}.
Taking the representative requirement $m_{U,D}\gtrsim 1~{\rm TeV}$ gives
\begin{eqnarray}
y_{U,D}\gtrsim \sqrt{2}\frac{1~{\rm TeV}}{246~{\rm GeV}}\simeq 6\;.
\label{eq:yukawa_bound}
\end{eqnarray}
Though this relation should be interpreted as a representative estimate rather
than a strict experimental lower bound,  an ordinary perturbative Higgs
realization is already driven to strong coupling.
Although introducing multiple Higgs doublets can reduce the magnitudes of the
individual Yukawa couplings $y_i$ needed to generate a given exotic-quark mass,
small individual couplings do not by themselves guarantee perturbative
unitarity\cite{Chanowitz:1978mv}. As emphasized by the partial-wave analysis of
Yukawa interactions in Ref.~\cite{Allwicher:2021rtd}, the relevant constraints
apply to the partial-wave scattering amplitudes, which can be enhanced by the
multiplicity of contributing states. In the present case, scalar exchange in the
high-energy, equal-helicity color-singlet quark--antiquark channel gives
$a_0=-N_c\sum_i|y_i|^2/(16\pi)$, with $N_c=3$. The usual tree-level perturbative
unitarity criterion is~\cite{Allwicher:2021rtd}
\begin{equation}
\left|\operatorname{Re}a_0\right|<\frac12.
\label{eq:quark_partial_wave_unitarity}
\end{equation}
For an exotic quark $Q$ whose mass arises entirely from Higgs doublets with VEVs
satisfying $\sum_i v_i^2=v^2$, the Cauchy--Schwarz inequality gives $m_Q^2\leq
v^2\sum_i|y_i|^2/2$. Combining this relation with the partial-wave bound yields
$m_Q<v\sqrt{4\pi/N_c}\simeq504~\mathrm{GeV}$ for $v=246~\mathrm{GeV}$.
Consequently, at fixed electroweak VEV, adding Higgs doublets cannot raise this
tree-level upper bound on \(m_Q\), even if the individual Yukawa couplings are
small.
The existence of new colored and charged fermions with masses generated by the
Higgs mechanism would also give nondecoupling loop contributions to $h\to gg$
and $h\to \gamma\gamma$ and cause problems compared with data for the
model~\cite{Djouadi:2005gi, He:2011ti, Eberhardt:2012gv,Bizot:2015zaa}.

We therefore regard Eq.~\eqref{eq:exotic_quark_yukawa} only as a useful
parametrization of the problem.
A viable model would require either a nonperturbative mass mechanism or new
dynamics that decouples the exotic-quark masses from the measured Higgs
couplings.
One possible, though not unique, completion is to introduce a second set of
fermions with opposite chiralities,  $Q'_R$, $U_R^c$, $D_R^c$, $E_R$, $N_R$.
Together with the original set, which we denote by $Q'_L$, $U_L^c$, $D_L^c$,
$E_L$, and $N_L$, these fields form a vectorlike exotic sector and allow
gauge-invariant Dirac mass terms. We have
\begin{equation}
-\mathcal{L} \supset M^{Q'} \bar Q'_L Q'_R + M^U \bar U^c_L U^c_R 
+ M^D \bar D^c_L D^c_R + M^E \bar E_L E_R 
+ M^{LR} \bar N_L N_R + {1\over 2} M^{LL}\bar N_L N^c_L 
+ {1\over 2} M^{RR}\bar N^c_R N_R + \mathrm{h.c.}
\end{equation}
Here the bare mass parameters are not fixed by the theory and they can be chosen
to be larger than the experimental bound. This will release the pressure on the
large size of Yukawa couplings discussed above which can instead remain small
enough to satisfy the relevant constraints. Meanwhile, the value of $q$ is not
fixed now. If $q\neq 1/2$, no renormalizable Yukawa couplings to the SM Higgs
$\Phi$ are allowed. If we identify the previous $N_1 = N_L$ and $N_2 = N^c_R$,
the previous discussion of the neutrino seesaw mechanism remains valid.

One may consider the above solution with vectorlike partners for the exotic
fermions.
However, anomaly cancellation is then automatic, which is not in the same spirit
as the chiral theory with which we started.
To address this issue, we propose another approach by invoking a technicolor
model~\cite{Weinberg:1975gm,Susskind:1978ms} for fermion mass generation.
We assume an exotic-quark set of technifermions transforming under a technicolor
group $SU(N)$:
$Q_L^{TC} = (U^{TC}_L, D^{TC}_L)^T: (N)(3,2,0)$,
$U_R^{TC}: (N)(3,1,+q)$, and $D_R^{TC}: (N)(3,1,-q)$.
The $N$ in the first parentheses represents the fundamental representation of
$SU(N)$ and is required to be even to prevent global Witten anomalies.
All other particles are singlets under this $SU(N)$ group.
For $q = 1/2$, the $SU(N)$ gauge interaction induces the condensates
$\langle \bar{U}^{TC}_L U^{TC}_R \rangle
=\langle \bar{D}^{TC}_L D^{TC}_R \rangle=\Lambda^3$,
which break the electroweak symmetry $SU(2)_L \times U(1)_Y$ down to $U(1)_{\rm
em}$.
Such a theory is compatible with anomaly cancellation~\cite{He:1989zb}.
Fermion masses are generated through four-fermion interactions involving exotic
quarks, which also contribute to the $W$ and $Z$ masses in the same way as the
Higgs VEV $v_H$:
\begin{equation}
\mathcal L_{4f}\supset{}
\frac{\lambda_U}{\Lambda^{\prime 2}}\,
\bigl(\bar{Q}'_{a} U \bigr)
\bigl(\bar{U}^{TC}_R Q_L^{TC\,a}\bigr)
+
\frac{\lambda_D}{\Lambda^{\prime 2}}\,
\bigl(\bar{Q}'_{a} D\bigr)
\bigl(\bar{D}^{TC}_R Q_L^{TC\,a}\bigr)
+\mathrm{h.c.}
\end{equation}
After condensation, the displayed operators generate exotic-quark
masses of magnitude $  m_{U,D}=|\lambda_{U,D}|\Lambda^3 /\Lambda^{\prime 2}.$
Here $\Lambda'$ is the scale suppressing the four-fermion operators. The
technicolor contribution to the weak gauge-boson masses, in the same way as the
Higgs VEV $v_H$, is determined by the total technipion decay constant $v_{TC}$.
The total effective VEV $v$ receives contributions from both $v_H$ and $v_{TC}$,
with $v^2 = (v_{H } )^2+ (v_{TC})^2$.
TeV-scale exotic-quark masses can be obtained with a small $v_{TC}$ depending on
the strong dynamics and large couplings.
For $v_{TC}\ll v$ and small mixing between the Higgs and technicolor scalar
states, this mechanism may avoid large changes to $h\to gg$ and
$h\to\gamma\gamma$.

The origin of exotic-quark masses remains the central open issue of this
construction and warrants further study. In the next section, we explore the
possibility that the exotic solution is associated not with the SM $\SU(2)_L$,
but with a new $\SU(2)_{L'}$ gauge group.
\\

\noindent{\bf A gauge group extension for the exotic solution.}
The nonstandard solution realizes a different hypercharge quantization, in which
nothing on the gauge side requires $q=1/2$.
For $q\neq1/2$, however, the exotic quarks cannot acquire Dirac masses in the SM
while preserving $\mathrm{U}(1)_{\rm em}$.
The lower component of an additional doublet $S_{\rm
SM}\sim(\mathbf{1},\mathbf{2},q)$ has charge $q-1/2$, so its VEV would break
$\mathrm{U}(1)_{\rm em}$. A chiral exotic sector with generic $q$ thus requires
the exotic doublets to transform under a gauge group other than $\SU(2)_L$.
Although this goes beyond the SM gauge structure assumed in the anomaly
analysis, it is a minimal and reasonable setup: one renormalizable completion
introduces $\SU(2)_{L'}$ acting only on the exotic sector~\cite{Hsieh:2010zr},
with gauge group
\begin{eqnarray}
G'=\SU(3)_C\times \SU(2)_L\times \SU(2)_{L'}\times \U(1)_X.
\end{eqnarray}
The particles in the nonstandard solution transform as
\begin{equation}
Q'\sim({\mathbf{3},\mathbf{1},\mathbf{2}},0),\quad
U^c\sim({\overline{\mathbf{3}},\mathbf{1},\mathbf{1}},-q),\quad 
 D^c\sim({\overline{\mathbf{3}},\mathbf{1},\mathbf{1}},q),\quad
E\sim({\mathbf{1},\mathbf{1},\mathbf{2}},0),\quad 
 N\sim({\mathbf{1},\mathbf{1},\mathbf{1}},0).
\label{eq:extended_exotic_representations}
\end{equation}
All SM multiplets are singlets under $\SU(2)_{L'}$. The scalar sector relevant
for symmetry breaking contains the SM Higgs
$\Phi\sim({\mathbf{1},\mathbf{2},\mathbf{1}},1/2)$ and an $\SU(2)_{L'}$ doublet
$S\sim({\mathbf{1},\mathbf{1},\mathbf{2}},q)$, where the last entry denotes the
$\U(1)_X$ charge.

The exotic quarks are $\SU(3)_C$ triplets, as needed for decays into SM quarks.
For anomaly counting, the right-handed fields may be replaced by their
left-handed conjugates; the field content still comprises 15 Weyl degrees of
freedom, $6+3+3+2+1$.
The ordering convention adopted above gives $q\geq0$. For $q>0$, the upper
component of $Q'$ has electric charge $+q$.
A single exotic family contains three $\SU(2)_{L'}$ doublets from the color
copies of $Q'$ and one from $E$, so the $\SU(2)_{L'}$ Witten anomaly is absent.
A massive exotic-lepton sector requires two copies of the exotic family, which
is the minimal choice. With two copies, the color-singlet $E$ doublets admit the
gauge-invariant vectorlike mass term $m_E E_{1}^T C i\sigma_2
E_{2}+\mathrm{h.c.}$.
This mass is independent of the $S$ vacuum expectation value, so $m_E$ need not
track the $\SU(2)_{L'}$-breaking scale $w$.

The most general renormalizable scalar potential is
\begin{equation}
V(\Phi,S)=-\mu_\Phi^2\Phi^\dagger\Phi-\mu_S^2S^\dagger S
+\lambda_\Phi(\Phi^\dagger\Phi)^2+\lambda_S(S^\dagger S)^2
+\lambda_{\Phi S}(\Phi^\dagger\Phi)(S^\dagger S).
\end{equation}
We take $\langle S\rangle=(0,w/\sqrt{2})^T$. The potential is bounded from below
for $\lambda_\Phi>0$, $\lambda_S>0$, and $\lambda_{\Phi
S}>-2\sqrt{\lambda_\Phi\lambda_S}$.
The Yukawa interactions
\begin{equation}
\mathcal L_Y\supset
    -y_D\overline{Q'}S D
    -y_U\overline{Q'}\widetilde S U
    +\mathrm{h.c.},
    \quad 
    \widetilde S=i\sigma_2S^\ast,
    \qquad
    m_{U,D}=\frac{y_{U,D}w}{\sqrt{2}},
\end{equation}
generate exotic-quark masses proportional to $w$, rather than to the SM Higgs
VEV. Symmetry breaking proceeds in two stages:
\begin{equation}
\SU(2)_L\times\SU(2)_{L'}\times\U(1)_X\to\SU(2)_L\times\U(1)_Y\to\U(1)_{\rm em}.
\end{equation}
 The residual $\U(1)_Y$ is identified with SM hypercharge.
The electric charge operator is
\begin{eqnarray}
Q_{\rm em}=T_L^3+Y,\qquad Y=2qT_{L'}^3+X .
\end{eqnarray}
Since the SM fields have $T_{L'}^3=0$, one has $X=Y$ in the SM sector.
With this definition of the electric charge operator, the upper and lower
components of the $\SU(2)_{L'}$ doublets $Q'$ and $E$ carry electric charges $q$
and $-q$, respectively.
For generic $q$, the exotic quarks cannot decay exclusively into SM particles
while conserving both color and electric charge.
The $\SU(2)_{L'}$ gauge symmetry prevents the exotic lepton doublet $E$ from
decaying into SM particles unless a scalar bidoublet connects the $\SU(2)_L$ and
$\SU(2)_{L'}$ sectors. Since no such bidoublet is included in our minimal setup,
the lightest exotic quark and lepton are stable.

We define the gauge couplings in the fermion sector by the covariant derivative
\begin{equation}
D_\mu=\partial_\mu
-i g_s G_\mu^A T_C^A
-i g_L W_\mu^a T_L^a
-i\widetilde g_{L'}W_\mu^{\prime a}T_{L'}^a
-i g_X X B_{X\mu}.
\end{equation}
Here $g_s$, $g_L$, $\widetilde g_{L'}$, and $g_X$ are the gauge couplings of
$\SU(3)_C$, $\SU(2)_L$, $\SU(2)_{L'}$, and $\U(1)_X$, respectively, while
$T_C^A$, $T_L^a$, $T_{L'}^a$, and $X$ are the corresponding generators. For
$q>0$, it is convenient to rescale the coupling and VEV by
$\widetilde g_{L'}\equiv 2qg_{L'},
w'\equiv 2qw$.
With only $\Phi$ and $S$ acquiring VEVs, there is no tree-level charged
$W$--$W'$ mixing; the charged gauge boson masses are $m_W=g_Lv/2$ and
$m_{W'}=g_{L'}w'/2$.
The leading tree-level electroweak correction arises in the neutral sector
because the two symmetry-breaking stages share the same Abelian gauge field. In
the convention above, the neutral gauge-boson mass terms are
\begin{equation}
\mathcal L_{\rm neutral}
=\frac{v^2}{8}\left(g_LW_\mu^3-g_XB_{X\mu}\right)^2
+\frac{w'^2}{8}\left(g_{L'}W_\mu^{\prime 3}-g_XB_{X\mu}\right)^2 .
\end{equation}
A convenient set of neutral fields consists of the hypercharge field $B_\mu$,
the SM-like field $Z_\mu^0$, the additional field $Z_\mu^{\prime0}$ orthogonal
to $B_\mu$, and the massless photon $A_\mu$:
\begin{equation}
\begin{aligned}
B_\mu&=\frac{g_{L'}B_{X\mu}+g_XW_\mu^{\prime 3}}{\sqrt{g_{L'}^2+g_X^2}},\qquad Z_\mu^{\prime 0}=\frac{g_{L'}W_\mu^{\prime
3}-g_XB_{X\mu}}{\sqrt{g_{L'}^2+g_X^2}},\\
Z_\mu^0&=\frac{g_LW_\mu^3-g_YB_\mu}{\sqrt{g_L^2+g_Y^2}},\qquad A_\mu=e\left(\frac{W_\mu^3}{g_L}
+\frac{B_{X\mu}}{g_X}
+\frac{W_\mu^{\prime 3}}{g_{L'}}\right).
\end{aligned}
\end{equation}
The corresponding hypercharge and electromagnetic couplings are
\begin{eqnarray}
g_Y\equiv\frac{g_Xg_{L'}}{\sqrt{g_{L'}^2+g_X^2}},
\qquad
\frac{1}{e^2}=\frac{1}{g_L^2}+\frac{1}{g_X^2}
+\frac{1}{g_{L'}^2}.
\end{eqnarray}
Neutral gauge-boson mixing lowers the physical $Z$-boson mass relative to its SM
tree-level value
\begin{equation}
m_Z^2\simeq
\frac{g_L^2+g_Y^2}{4}v^2
\left[
1-\left(\frac{g_Y}{g_{L'}}\right)^4\frac{v^2}{w'^2}
\right].
\end{equation}
Formally, $m_W$ is not shifted because $W$--$W'$ mixing is absent. We note that,
in the SM, the following relation holds
\begin{equation}
(m_W^{\rm SM})^2 =
\frac{m_Z^2}{2}
\left[
1+
\sqrt{
1-\frac{4\pi\alpha_{\rm em}}
{\sqrt{2}G_F m_Z^2}
}
\right].
\end{equation}
The shift in the $Z$-boson mass modifies $m_W$ relative to   $m_W^{\rm SM}$ by
\begin{equation}
\delta m_W^2
\equiv \frac{g_L^2 v^2 }{4 } -
(m_W^{\rm SM})^2
\simeq
\frac{(m_W^{\rm SM})^4}
{2(m_W^{\rm SM})^2-m_Z^2}
\left(\frac{g_Y}{g_{L'}}\right)^4
\frac{v^2}{w'^2}.
\end{equation}
Here, $m_W^{\rm SM}$ is determined from the inputs of $m_Z$, $G_F$ and
$\alpha_{\rm em}$.
The current direct measurement and SM prediction give $m_W^{\rm
exp}=80.3625\pm0.0077~{\rm GeV}$ and $m_W^{\rm SM}=80.357\pm0.006~{\rm GeV}$,
respectively~\cite{ParticleDataGroup:2026mpi}. Using  $\alpha_{\rm
em}^{-1}(m_Z^2)=128.947\pm0.013$~\cite{Davier:2019can},
we obtain the following constrained at the 95\% confidence level as
\begin{equation}
w'\gtrsim11.9~{\rm TeV}
\left(\frac{g_Y}{g_{L'}}\right)^2.
\end{equation}
Since $m_{U,D}=y_{U,D}w/\sqrt{2}$, order-one Yukawa couplings and weak-sized
$g_{L'}$ naturally place the exotic quarks near the $10\,\mathrm{TeV}$ scale.

Because \(N\) remains a gauge singlet and the \(E\) doublets admit the same
antisymmetric mass term under \(\SU(2)_{L'}\), the preceding minimal seesaw and
exotic lepton mass constructions remain valid, although the electric charges are
now \(\pm q\).

In this model, there are charged particles with electric charge $\pm q$, and
they are constrained by collider and cosmic-ray searches.
For multiply charged particles with $2\leq q\leq8$, recasts of open-production
searches give approximate lower mass limits of $1.2~{\rm TeV}$ for color-triplet
fermions, and around $0.8~{\rm TeV}$ for color-singlet
fermions~\cite{Jager:2018ecz}. For stable color-singlet fermions with fractional
charges $1/3\leq q <1$, CMS excludes masses ranging from $60$ to $640
\mathrm{GeV}$, depending on $q$~\cite{CMS:2024eyx}.
Cosmic-ray searches provide complementary constraints, especially for
fractionally charged particles (FCPs, $0<q<1$). Ref.~\cite{SuperCDMS:2020hcc}
summarizes constraints on the FCP mass--charge parameter space from
astrophysical observations and direct laboratory experiments. Depending on the
values of $q$ and the masses of the exotic particles in our model, different
constraints apply
\cite{MACRO:2000bht,CDMS:2014ane,Majorana:2018gib,TEXONO:2018nir,SuperCDMS:2020hcc,Plestid:2020kdm}.
The strongest mass constraint comes from CDMSlite, which covers
$5~\mathrm{MeV}\lesssim m_E\lesssim100~\mathrm{TeV}$ and gives the strongest
direct-detection limits for $q\leq1/160$.
Since both \(q\) and the exotic-particle masses are free parameters in our
model, they can be safely chosen outside the excluded regions, for example
$q>1/160$ and masses of order a TeV or larger. The above constraints are also
relevant to the case discussed in the previous section where the stable exotic
particles carry electric charges $\pm1/2$. The discovery of some of these exotic
particles above the current collider bounds may provide evidence for these
models.

If the new $\SU(2)$ is instead viewed as a right-handed extension, the exotic
doublets $Q'_R$, $E_R$, and the singlet $N_R$ are right-handed and the singlets
$U_L\sim(\mathbf3,\mathbf1,\mathbf1,q)$,
$D_L\sim(\mathbf3,\mathbf1,\mathbf1,-q)$ are left-handed, with anomaly
cancellation unchanged.
For $q=2/3$, $U_L$ carries the quantum numbers of $u_R$, so the mixing
$-\mu\overline U_Lu_R$ is allowed alongside $-y'_U\overline U_L\widetilde
S^\dagger Q'_R$. With $Q'_R=(U'_R,D'_R)^T$, $m_u^{(0)}=y_uv/\sqrt2$, and
$M_U=y'_Uw/\sqrt2$, the first-generation up-quark mass matrix is
\begin{equation}
-\mathcal L_m\supset
\begin{pmatrix}\overline u_L&\overline U_L\end{pmatrix}
\begin{pmatrix}m_u^{(0)}&0\\ \mu&M_U\end{pmatrix}
\begin{pmatrix}u_R\\ U'_R\end{pmatrix}
+\mathrm{h.c.},
\label{eq:up_mixing_mass_matrix}
\end{equation}
which gives $m_u\simeq|m_u^{(0)}|/\sqrt{1+|\mu/M_U|^2}$ and $M_{\rm
heavy}\simeq\sqrt{|\mu|^2+|M_U|^2}$. For $|\mu|\gg|M_U|$, this reduces to
$m_u\simeq|m_u^{(0)}M_U/\mu|$, a seesaw in the quark sector, so a large $\mu$
suppresses the up-quark mass.

Such gauge invariant mixing permits the corresponding singlet exotic quark to
decay. Taking the top quark as an example, the similar mixing term $-\mu_U
\overline{U}_L t_R$ is allowed.
The leading expansion is controlled by $m_t^{(0)}/\sqrt{|M_U|^2 + |\mu_U|^2}$.
This mixing affects Cabibbo-Kobayashi-Maskawa (CKM) physics. The left-handed
mixing angle is approximately
\begin{eqnarray}
    |\theta_L| \simeq \frac{m_t^{(0)} |\mu_U|}{|\mu_U|^2 + |M_U|^2}.
\end{eqnarray}
Therefore the observed charged-current couplings are approximately $V_{t j}
\simeq (1-\theta_L^2/2) V_{t j}^{(0)},\allowbreak V_{T j} \simeq \theta_L V_{t j}^{(0)}$
for $j = d,s,b$.
The full four quark charged-current matrix remains unitary, but its $3 \times 3$
SM sub-block is not exactly unitary:
\begin{eqnarray}
\label{eq:ckm-nonclosure}
V_{ud}V_{ub}^\ast+V_{cd}V_{cb}^\ast+V_{td}V_{tb}^\ast=-V_{Td}V_{Tb}^\ast .
\end{eqnarray}
Therefore the usual CKM unitarity triangle does not close exactly; the leading
deviation is of order $|\theta_L|^2$ and is small in the decoupling limit.
\\

\noindent{\bf Summary.}
We have analyzed gauge anomaly cancellation for a Standard-Model-like set of 15
fermions with fixed non-Abelian representations and arbitrary hypercharges,
focusing on the nonstandard solution.
This solution contains exotic doublets and singlets with zero hypercharge,
together with color anti-triplets carrying hypercharges $-q$ and $q$. The
neutral singlets can facilitate a type-I seesaw mechanism to generate neutrino
masses.
The antisymmetric lepton mass matrix requires an even number of copies for all
exotic lepton doublets to acquire nonzero masses. With two copies, the neutrino
mass matrix generated by the type-I seesaw has rank two, leaving one active
neutrino massless. For $0\nu\beta\beta$ decay, our model predicts the effective
Majorana mass $m_{\beta\beta}=1.17\text{--}4.07~\mathrm{meV}$ for NO and
$15.85\text{--}48.94~\mathrm{meV}$ for IO. Future searches can probe most of the
IO range, while the NO range remains beyond reach.

If the exotic quarks get their masses from the SM Higgs, gauge invariance fixes
$q=1/2$. A single exotic quark cannot decay into SM particles, so the lightest
one is stable. Representative TeV-scale mass limits require $y_{U,D}\gtrsim 6$,
which is in the strong-coupling regime. These particles also give nondecoupling
corrections to $h\to gg$ and $h\to\gamma\gamma$, so Higgs data strongly restrict
this minimal case. Opposite chirality partners allow gauge-invariant Dirac
masses, but make the fermion content vectorlike and anomaly cancellation
automatic.
Alternatively, a technicolor approach can dynamically generate large
exotic-quark masses. In this setup, technifermion condensates break the
electroweak gauge symmetry without Higgs Yukawa coupling. The exotic lepton
doublet contains stable particles with electric charges $\pm1/2$, providing a
potential smoking-gun signature of the model.
Alternatively, a separate $\SU(2)_{L'}$ can generate the exotic-quark masses at
a new scale $w$ rather than the SM Higgs vacuum expectation value. The $Z$--$Z'$
mixing bound, $2 q w\gtrsim 11.9 \,\mathrm{TeV}\,(g_Y/g_{L'})^2$, still permits
exotic-quark masses from several TeV to about $10~\mathrm{TeV}$ for order-one
Yukawa couplings.
For $q=2/3$, mixing with an opposite-chirality exotic family can suppress the
up-quark mass and affect CKM physics.
In our model, though some exotic charged particles are stable and carry electric
charges $\pm q$ or $\pm 1/2$, they can still be consistent with collider and
cosmic-ray constraints.
\\

\noindent{
\bf Acknowledgments
}
This work is supported in part by the National Key Research and Development
Program of China under Grant
No. 2020YFC2201501, by the Fundamental Research
Funds for the Central Universities, by the National Natural Science Foundation
of P.R. China (Nos. 12090064, 12375088, 12575096 and W2441004).

\bibliography{ref}

@article{Adler:1969gk,
    author = "Adler, Stephen L.",
    title = "{Axial vector vertex in spinor electrodynamics}",
    doi = "10.1103/PhysRev.177.2426",
    journal = "Phys. Rev.",
    volume = "177",
    pages = "2426--2438",
    year = "1969"
}

@article{Bell:1969ts,
    author = "Bell, J. S. and Jackiw, R.",
    title = "{A PCAC puzzle: $\pi^0 \to \gamma \gamma$ in the $\sigma$ model}",
    doi = "10.1007/BF02823296",
    journal = "Nuovo Cim. A",
    volume = "60",
    pages = "47--61",
    year = "1969"
}

@article{Bizot:2015zaa,
    author = "Bizot, Nicolas and Frigerio, Michele",
    title = "{Fermionic extensions of the Standard Model in light of the Higgs couplings}",
    eprint = "1508.01645",
    archivePrefix = "arXiv",
    primaryClass = "hep-ph",
    doi = "10.1007/JHEP01(2016)036",
    journal = "JHEP",
    volume = "01",
    number = "2016",
    pages = "036",
    year = "2016"
}

@article{Bouchiat:1972iq,
    author = "Bouchiat, C. and Iliopoulos, J. and Meyer, P.",
    title = "{An Anomaly Free Version of Weinberg's Model}",
    doi = "10.1016/0370-2693(72)90532-1",
    journal = "Phys. Lett. B",
    volume = "38",
    pages = "519--523",
    year = "1972"
}

@article{tHooft:1971qjg,
    author = "'t Hooft, Gerard",
    editor = "Taylor, J. C.",
    title = "{Renormalizable Lagrangians for Massive Yang-Mills Fields}",
    doi = "10.1016/0550-3213(71)90139-8",
    journal = "Nucl. Phys. B",
    volume = "35",
    pages = "167--188",
    year = "1971"
}

@article{Gross:1972pv,
    author = "Gross, David J. and Jackiw, R.",
    title = "{Effect of anomalies on quasirenormalizable theories}",
    doi = "10.1103/PhysRevD.6.477",
    journal = "Phys. Rev. D",
    volume = "6",
    pages = "477--493",
    year = "1972"
}

@article{Geng:1989tcu,
    author = "Geng, C. Q. and Marshak, R. E.",
    title = "{Uniqueness of Quark and Lepton Representations in the Standard Model From the Anomalies Viewpoint}",
    reportNumber = "VPI-IHEP-88-5",
    doi = "10.1103/PhysRevD.39.693",
    journal = "Phys. Rev. D",
    volume = "39",
    pages = "693",
    year = "1989"
}

@article{Minahan:1989vd,
  title = {Comment on Anomaly Cancellation in the Standard Model},
  author = {Minahan, J. A. and Ramond, P. and Warner, R. C.},
  journal = {Phys. Rev. D},
  volume = {41},
  number = {2},
  pages = {715--716},
  year = {1990},
  doi = {10.1103/PhysRevD.41.715},
  url = {https://link.aps.org/doi/10.1103/PhysRevD.41.715},
}

@article{Foot:1992ui,
    author = "Foot, Robert and Lew, H. and Volkas, R. R.",
    title = "{Electric charge quantization}",
    eprint = "hep-ph/9209259",
    archivePrefix = "arXiv",
    reportNumber = "UM-P-92-52, OZ-92-19, SHEP-91-92",
    doi = "10.1088/0954-3899/19/3/005",
    journal = "J. Phys. G",
    volume = "19",
    pages = "361--372",
    year = "1993",
    note = "[Erratum: J.Phys.G 19, 1067 (1993)]"
}

@article{Babu:1989ex,
    author = "Babu, K. S. and Mohapatra, Rabindra N.",
    title = "{Quantization of Electric Charge From Anomaly Constraints and a Majorana Neutrino}",
    reportNumber = "MDDP-PP-90-011",
    doi = "10.1103/PhysRevD.41.271",
    journal = "Phys. Rev. D",
    volume = "41",
    pages = "271",
    year = "1990"
}

@article{Geng:1990nh,
  title = {Reply to ``Comment on Anomaly Cancellation in the Standard Model''},
  author = {Geng, C. Q. and Marshak, R. E.},
  journal = {Phys. Rev. D},
  volume = {41},
  number = {2},
  pages = {717--718},
  year = {1990},
  doi = {10.1103/PhysRevD.41.717},
  url = {https://link.aps.org/doi/10.1103/PhysRevD.41.717},
}

@misc{NuFIT:6.1,
    author = "Esteban, Ivan and Gonzalez-Garcia, M. C. and Maltoni, Michele and Martinez-Soler, Ivan and Pinheiro, Joao Paulo and Schwetz, Thomas",
    title = "{NuFIT 6.1 (2025)}",
    year = "2025",
    howpublished = "\url{https://www.nu-fit.org/?q=node/309}",
    note = "Three-neutrino fit based on data available in November 2025"
}

@article{Frampton:2002qc,
    author = "Frampton, P. H. and Glashow, S. L. and Yanagida, T.",
    title = "{Cosmological sign of neutrino CP violation}",
    eprint = "hep-ph/0208157",
    archivePrefix = "arXiv",
    reportNumber = "CERN-TH-2002-193",
    doi = "10.1016/S0370-2693(02)02853-8",
    journal = "Phys. Lett. B",
    volume = "548",
    pages = "119--121",
    year = "2002"
}

@article{He:1990me,
    author = "He, X. G. and Joshi, Girish C. and Volkas, R. R.",
    title = "{Constraints from anomaly cancellation on strong, weak, and electromagnetic interactions}",
    doi = "10.1103/PhysRevD.41.278",
    journal = "Phys. Rev. D",
    volume = "41",
    pages = "278--280",
    year = "1990"
}

@article{Eberhardt:2012gv,
    author = "Eberhardt, Otto and Herbert, Geoffrey and Lacker, Heiko and Lenz, Alexander and Menzel, Andreas and Nierste, Ulrich and Wiebusch, Martin",
    title = "{Impact of a Higgs boson at a mass of 126 GeV on the standard model with three and four fermion generations}",
    eprint = "1209.1101",
    archivePrefix = "arXiv",
    primaryClass = "hep-ph",
    reportNumber = "TTP12-034",
    doi = "10.1103/PhysRevLett.109.241802",
    journal = "Phys. Rev. Lett.",
    volume = "109",
    pages = "241802",
    year = "2012"
}

@article{Delbourgo:1972xb,
    author = "Delbourgo, Robert and Salam, Abdus",
    title = "{The gravitational correction to pcac}",
    reportNumber = "ICTP-71-24",
    doi = "10.1016/0370-2693(72)90825-8",
    journal = "Phys. Lett. B",
    volume = "40",
    pages = "381--382",
    year = "1972"
}

@article{Alvarez-Gaume:1983ihn,
    author = "Alvarez-Gaume, Luis and Witten, Edward",
    editor = "Salam, A. and Sezgin, E.",
    title = "{Gravitational Anomalies}",
    reportNumber = "HUTP-83/A039",
    doi = "10.1016/0550-3213(84)90066-X",
    journal = "Nucl. Phys. B",
    volume = "234",
    pages = "269",
    year = "1984"
}

@article{Witten:1982fp,
    author = "Witten, Edward",
    editor = "Shifman, Mikhail A.",
    title = "{An SU(2) Anomaly}",
    doi = "10.1016/0370-2693(82)90728-6",
    journal = "Phys. Lett. B",
    volume = "117",
    pages = "324--328",
    year = "1982"
}

@article{CMS:2024eyx,
    author = "Hayrapetyan, Aram and others",
    collaboration = "CMS",
    title = "{Search for Fractionally Charged Particles in Proton-Proton Collisions at s=13{\,}{\,}TeV}",
    eprint = "2402.09932",
    archivePrefix = "arXiv",
    primaryClass = "hep-ex",
    reportNumber = "CMS-EXO-19-006, CERN-EP-2024-002",
    doi = "10.1103/PhysRevLett.134.131802",
    journal = "Phys. Rev. Lett.",
    volume = "134",
    number = "13",
    pages = "131802",
    year = "2025"
}

@article{He:2011ti,
    author = "He, Xiao-Gang and Valencia, German",
    title = "{An extended scalar sector to address the tension between a fourth generation and Higgs searches at the LHC}",
    eprint = "1108.0222",
    archivePrefix = "arXiv",
    primaryClass = "hep-ph",
    doi = "10.1016/j.physletb.2011.12.063",
    journal = "Phys. Lett. B",
    volume = "707",
    pages = "381--384",
    year = "2012"
}

@article{SuperCDMS:2020hcc,
    author = "Alkhatib, I. and others",
    collaboration = "SuperCDMS",
    title = "{Constraints on Lightly Ionizing Particles from CDMSlite}",
    eprint = "2011.09183",
    archivePrefix = "arXiv",
    primaryClass = "hep-ex",
    doi = "10.1103/PhysRevLett.127.081802",
    journal = "Phys. Rev. Lett.",
    volume = "127",
    number = "8",
    pages = "081802",
    year = "2021"
}

@article{Plestid:2020kdm,
    author = "Plestid, Ryan and Takhistov, Volodymyr and Tsai, Yu-Dai and Bringmann, Torsten and Kusenko, Alexander and Pospelov, Maxim",
    title = "{New Constraints on Millicharged Particles from Cosmic-ray Production}",
    eprint = "2002.11732",
    archivePrefix = "arXiv",
    primaryClass = "hep-ph",
    reportNumber = "FERMILAB-PUB-20-044-A-T, INT-PUB-20-004, IPMU20-0015",
    doi = "10.1103/PhysRevD.102.115032",
    journal = "Phys. Rev. D",
    volume = "102",
    pages = "115032",
    year = "2020"
}

@article{MACRO:2000bht,
    author = "Ambrosio, M. and others",
    collaboration = "MACRO",
    title = "{A Search for lightly ionizing particles with the MACRO detector}",
    eprint = "hep-ex/0002029",
    archivePrefix = "arXiv",
    doi = "10.1103/PhysRevD.62.052003",
    journal = "Phys. Rev. D",
    volume = "62",
    pages = "052003",
    year = "2000"
}

@article{CDMS:2014ane,
    author = "Agnese, R. and others",
    collaboration = "CDMS",
    title = "{First Direct Limits on Lightly Ionizing Particles with Electric Charge Less Than $e/6$}",
    eprint = "1409.3270",
    archivePrefix = "arXiv",
    primaryClass = "hep-ex",
    reportNumber = "FERMILAB-PUB-14-418-AE",
    doi = "10.1103/PhysRevLett.114.111302",
    journal = "Phys. Rev. Lett.",
    volume = "114",
    number = "11",
    pages = "111302",
    year = "2015"
}

@article{Majorana:2018gib,
    author = "Alvis, S. I. and others",
    collaboration = "Majorana",
    title = "{First Limit on the Direct Detection of Lightly Ionizing Particles for Electric Charge as Low as e/1000 with the Majorana Demonstrator}",
    eprint = "1801.10145",
    archivePrefix = "arXiv",
    primaryClass = "hep-ex",
    doi = "10.1103/PhysRevLett.120.211804",
    journal = "Phys. Rev. Lett.",
    volume = "120",
    number = "21",
    pages = "211804",
    year = "2018"
}

@article{TEXONO:2018nir,
    author = "Singh, L. and others",
    collaboration = "TEXONO",
    title = "{Constraints on millicharged particles with low threshold germanium detectors at Kuo-Sheng Reactor Neutrino Laboratory}",
    eprint = "1808.02719",
    archivePrefix = "arXiv",
    primaryClass = "hep-ph",
    doi = "10.1103/PhysRevD.99.032009",
    journal = "Phys. Rev. D",
    volume = "99",
    number = "3",
    pages = "032009",
    year = "2019"
}

@article{Jager:2018ecz,
    author = {J{\"a}ger, Sebastian and Kvedarait{\.{e}}, Sandra and Perez, Gilad and Savoray, Inbar},
    title = "{Bounds and prospects for stable multiply charged particles at the LHC}",
    eprint = "1812.03182",
    archivePrefix = "arXiv",
    primaryClass = "hep-ph",
    doi = "10.1007/JHEP04(2019)041",
    journal = "JHEP",
    volume = "04",
    number = "2019",
    pages = "041",
    year = "2019"
}

@article{ParticleDataGroup:2026mpi,
    author = "Takahashi, F. and others",
    collaboration = "Particle Data Group",
    title = "{Review of Particle Physics}",
    doi = "10.1142/S0217751X26300115",
    journal = "Int. J. Mod. Phys. A",
    volume = "41",
    pages = "2630011",
    year = "2026"
}

@article{Djouadi:2005gi,
    author = "Djouadi, Abdelhak",
    title = "{The Anatomy of electro-weak symmetry breaking. I: The Higgs boson in the standard model}",
    eprint = "hep-ph/0503172",
    archivePrefix = "arXiv",
    reportNumber = "LPT-ORSAY-05-17",
    doi = "10.1016/j.physrep.2007.10.004",
    journal = "Phys. Rept.",
    volume = "457",
    pages = "1--216",
    year = "2008"
}

@article{Minkowski:1977sc,
    author = "Minkowski, Peter",
    title = "{$\mu \to e\gamma$ at a Rate of One Out of $10^{9}$ Muon Decays?}",
    reportNumber = "Print-77-0182 (BERN)",
    doi = "10.1016/0370-2693(77)90435-X",
    journal = "Phys. Lett. B",
    volume = "67",
    pages = "421--428",
    year = "1977"
}

@article{Yanagida:1979as,
    author = "Yanagida, Tsutomu",
    editor = "Sawada, Osamu and Sugamoto, Akio",
    title = "{Horizontal gauge symmetry and masses of neutrinos}",
    reportNumber = "KEK-79-18-95",
    journal = "Conf. Proc. C",
    volume = "7902131",
    pages = "95--99",
    year = "1979"
}

@article{Gell-Mann:1979vob,
    author = "Gell-Mann, Murray and Ramond, Pierre and Slansky, Richard",
    title = "{Complex Spinors and Unified Theories}",
    eprint = "1306.4669",
    archivePrefix = "arXiv",
    primaryClass = "hep-th",
    reportNumber = "PRINT-80-0576",
    journal = "Conf. Proc. C",
    volume = "790927",
    pages = "315--321",
    year = "1979"
}

@article{Mohapatra:1979ia,
    author = "Mohapatra, Rabindra N. and Senjanovic, Goran",
    title = "{Neutrino Mass and Spontaneous Parity Nonconservation}",
    reportNumber = "MDDP-TR-80-060, MDDP-PP-80-105, CCNY-HEP-79-10",
    doi = "10.1103/PhysRevLett.44.912",
    journal = "Phys. Rev. Lett.",
    volume = "44",
    pages = "912",
    year = "1980"
}

@article{Ibarra:2003up,
    author = "Ibarra, A. and Ross, Graham G.",
    title = "{Neutrino phenomenology: The Case of two right-handed neutrinos}",
    eprint = "hep-ph/0312138",
    archivePrefix = "arXiv",
    reportNumber = "CERN-TH-2003-294, OUTP-0333P",
    doi = "10.1016/j.physletb.2004.04.037",
    journal = "Phys. Lett. B",
    volume = "591",
    pages = "285--296",
    year = "2004"
}

@article{Davidson:2006tg,
    author = "Davidson, Sacha and Isidori, Gino and Strumia, Alessandro",
    title = "{The Smallest neutrino mass}",
    eprint = "hep-ph/0611389",
    archivePrefix = "arXiv",
    reportNumber = "IFUP-TH-06-23",
    doi = "10.1016/j.physletb.2007.01.015",
    journal = "Phys. Lett. B",
    volume = "646",
    pages = "100--104",
    year = "2007"
}

@article{Furry:1939qr,
    author = "Furry, W. H.",
    title = "{On transition probabilities in double beta-disintegration}",
    doi = "10.1103/PhysRev.56.1184",
    journal = "Phys. Rev.",
    volume = "56",
    pages = "1184--1193",
    year = "1939"
}

@article{Schechter:1981bd,
    author = "Schechter, J. and Valle, J. W. F.",
    title = "{Neutrinoless Double beta Decay in SU(2) x U(1) Theories}",
    reportNumber = "SU-4217-213, COO-3533-213",
    doi = "10.1103/PhysRevD.25.2951",
    journal = "Phys. Rev. D",
    volume = "25",
    pages = "2951",
    year = "1982"
}

@article{Hsieh:2010zr,
    author = "Hsieh, Ken and Schmitz, Kai and Yu, Jiang-Hao and Yuan, C. -P.",
    title = "{Global Analysis of General SU(2) x SU(2) x U(1) Models with Precision Data}",
    eprint = "1003.3482",
    archivePrefix = "arXiv",
    primaryClass = "hep-ph",
    reportNumber = "MSUHEP-091123, DESY-09-205",
    doi = "10.1103/PhysRevD.82.035011",
    journal = "Phys. Rev. D",
    volume = "82",
    pages = "035011",
    year = "2010"
}

@article{Esteban:2024eli,
    author = "Esteban, Ivan and Gonzalez-Garcia, M. C. and Maltoni, Michele and Martinez-Soler, Ivan and Pinheiro, Jo{\~a}o Paulo and Schwetz, Thomas",
    title = "{NuFit-6.0: updated global analysis of three-flavor neutrino oscillations}",
    eprint = "2410.05380",
    archivePrefix = "arXiv",
    primaryClass = "hep-ph",
    reportNumber = "IFT-UAM/CSIC-24-140, YITP-SB-2024-24, IPPP/24/64, IPPP/24/64, IFT-UAM/CSIC-24-140, YITP-SB-2024-24",
    doi = "10.1007/JHEP12(2024)216",
    journal = "JHEP",
    volume = "12",
    number = "2024",
    pages = "216",
    year = "2024"
}

@article{JUNO:2025gmd,
    author = "Abusleme, Angel and others",
    collaboration = "JUNO",
    title = "{Measurement of reactor neutrino oscillation with the first JUNO data}",
    eprint = "2511.14593",
    archivePrefix = "arXiv",
    primaryClass = "hep-ex",
    doi = "10.1038/s41586-026-10538-z",
    journal = "Nature",
    volume = "654",
    number = "8118",
    pages = "343--348",
    year = "2026"
}

@article{Nakamura:2020szx,
    author = "Nakamura, R. and Sambonsugi, H. and Shiraishi, K. and Wada, Y.",
    editor = "Nakahata, Masayuki",
    title = "{Research and development toward KamLAND2-Zen}",
    doi = "10.1088/1742-6596/1468/1/012256",
    journal = "J. Phys. Conf. Ser.",
    volume = "1468",
    number = "1",
    pages = "012256",
    year = "2020"
}

@article{KamLAND-Zen:2024eml,
    author = "Abe, S. and others",
    collaboration = "KamLAND-Zen",
    title = "{Search for Majorana Neutrinos with the Complete KamLAND-Zen Dataset}",
    eprint = "2406.11438",
    archivePrefix = "arXiv",
    primaryClass = "hep-ex",
    doi = "10.1103/jkf6-48j8",
    journal = "Phys. Rev. Lett.",
    volume = "135",
    number = "26",
    pages = "262501",
    year = "2025"
}

@article{Allwicher:2021rtd,
    author = "Allwicher, Lukas and Arnan, Pere and Barducci, Daniele and Nardecchia, Marco",
    title = "{Perturbative unitarity constraints on generic Yukawa interactions}",
    eprint = "2108.00013",
    archivePrefix = "arXiv",
    primaryClass = "hep-ph",
    doi = "10.1007/JHEP10(2021)129",
    journal = "JHEP",
    volume = "10",
    number = "2021",
    pages = "129",
    year = "2021"
}

@article{Esteban:2026phq,
    author = "Esteban, Ivan and Gonzalez-Garcia, M. C. and Maltoni, Michele and Martinez-Soler, Ivan and Pinheiro, Joao Paulo and Schwetz, Thomas",
    title = "{Lessons from the first JUNO results}",
    eprint = "2601.09791",
    archivePrefix = "arXiv",
    primaryClass = "hep-ph",
    reportNumber = "IFT-UAM/CSIC-26-3, IPPP/26/03, YITP-SB-2026-02",
    doi = "10.1007/JHEP04(2026)089",
    journal = "JHEP",
    volume = "04",
    number = "2026",
    pages = "089",
    year = "2026"
}

@article{Davier:2019can,
    author = "Davier, M. and Hoecker, A. and Malaescu, B. and Zhang, Z.",
    title = "{A new evaluation of the hadronic vacuum polarisation contributions to the muon anomalous magnetic moment and to $\mathbf{\boldsymbol\alpha(m_Z^2)}$}",
    eprint = "1908.00921",
    archivePrefix = "arXiv",
    primaryClass = "hep-ph",
    doi = "10.1140/epjc/s10052-020-7792-2",
    journal = "Eur. Phys. J. C",
    volume = "80",
    number = "3",
    pages = "241",
    year = "2020",
    note = "[Erratum: Eur.Phys.J.C 80, 410 (2020)]"
}

@article{Dev:2026ddq,
    author = "Dev, P. S. Bhupal and Gehrlein, Julia and Sengupta, Amartya and Soni, Amarjit",
    title = "{Minimal dark SU(2) origin of a massless Dirac neutrino}",
    eprint = "2605.28923",
    archivePrefix = "arXiv",
    primaryClass = "hep-ph",
    doi = "10.1103/h3y7-qzyr",
    journal = "Phys. Rev. D",
    volume = "114",
    number = "3",
    pages = "L031702",
    year = "2026"
}

@article{Chanowitz:1978mv,
    author = "Chanowitz, Michael S. and Furman, M. A. and Hinchliffe, I.",
    title = "{Weak Interactions of Ultraheavy Fermions. 2.}",
    reportNumber = "LBL-8270",
    doi = "10.1016/0550-3213(79)90606-0",
    journal = "Nucl. Phys. B",
    volume = "153",
    pages = "402--430",
    year = "1979"
}

@article{He:1989zb,
    author = "He, X. G. and Joshi, Girish C. and Lew, H. and McKellar, B. H. J. and Volkas, R. R.",
    title = "{Charge Quantization in Supersymmetric, Technicolor and Composite Models}",
    reportNumber = "UM-P-89/82, OZ-P-89/27",
    doi = "10.1103/PhysRevD.40.3140",
    journal = "Phys. Rev. D",
    volume = "40",
    pages = "3140",
    year = "1989"
}

@article{Weinberg:1975gm,
    author = "Weinberg, Steven",
    title = "{Implications of Dynamical Symmetry Breaking}",
    reportNumber = "PRINT-75-0804 (HARVARD)",
    doi = "10.1103/PhysRevD.19.1277",
    journal = "Phys. Rev. D",
    volume = "13",
    pages = "974--996",
    year = "1976",
    note = "[Addendum: Phys.Rev.D 19, 1277--1280 (1979)]"
}

@article{Susskind:1978ms,
    author = "Susskind, Leonard",
    title = "{Dynamics of Spontaneous Symmetry Breaking in the Weinberg-Salam Theory}",
    reportNumber = "SLAC-PUB-2142",
    doi = "10.1103/PhysRevD.20.2619",
    journal = "Phys. Rev. D",
    volume = "20",
    pages = "2619--2625",
    year = "1979"
}

\end{document}